\documentclass[12pt]{article}
\usepackage{geometry}
\usepackage{graphicx}
\usepackage{booktabs} 
\usepackage{array} 
\usepackage{paralist} 
\usepackage{verbatim} 
\usepackage{subfig} 
\usepackage{changes}
\usepackage{amsmath}
\usepackage{bbm}
\usepackage{dsfont}
\usepackage{amsthm,amsfonts,amssymb}
\usepackage{mathtools}
\usepackage{ulem}
\usepackage[active]{srcltx}
\usepackage{hyperref}
\hypersetup{colorlinks=true,pdfborder={0 0 0}}

\usepackage[affil-it]{authblk}
\usepackage{tikz}
 \usepackage{tikz-3dplot}
 \usepackage{stackengine}
\usepackage{fancyhdr} 
\usepackage{sectsty}
\allsectionsfont{\sffamily\mdseries\upshape} 

\usepackage[nottoc,notlof,notlot]{tocbibind} 
\usepackage[titles,subfigure]{tocloft} 

\def \be{\begin{equation}}
\def \ee{\end{equation}}
\def\id{\mathds{1}}

\newcommand{\red}[1]{{\color{red} #1}}

\newcommand{\bra}[1]{\left\langle #1 \right|}
\newcommand{\braket}[2]{\langle #1 | #2 \rangle}
\newcommand{\ket}[1]{\left| #1 \right\rangle}

\newcommand{\mbf}[1]{\mathbf{ #1} }
\newcommand{\average}[1]{\mathbb{E}\left( #1\right)}

\DeclareMathAlphabet\mathbfcal{OMS}{cmsy}{b}{n}
\numberwithin{equation}{section}
\newtheorem{exa}{Example}[section]
\newtheorem{rem}{Remark}[section]

\title{An elementary introduction to the geometry of  {quantum states} with pictures
}
\author {J. Avron, O. Kenneth
}
\affil{Dept. of Physics, Technion, Israel}
\date{\today}
\begin{document}
\maketitle
\begin{abstract}
This is a review of the  geometry  of quantum states using elementary methods and pictures.  {Quantum states are represented by a convex body, often in  high dimensions. In the case of $n$-qubits, the dimension is exponentially large in $n$. The space of states can be visualized, to some extent, by its simple cross sections: Regular simplexes, balls and hyper-octahedra\footnote{\label{fn:octa}Also known as cross polytope,  orthoplex, and  co-cube}. When the dimension gets large there is a precise sense in which the space of states resembles, almost in every direction, a ball. The ball turns out to be a ball of rather low purity states.}  We also address some of the corresponding, but harder, geometric properties of separable and entangled states {and entanglement witnesses.}.
\end{abstract}
\begin{flushright}
{\it ``All convex bodies behave a bit like Euclidean balls." \\
\hfill Keith Ball}
\end{flushright}

\tableofcontents
\section{Introduction}
\subsection{The geometry of quantum states}
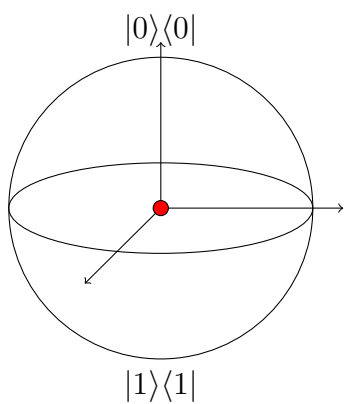
\begin{figure}[ht!]
\centering{}
\begin{tikzpicture}[scale=2]
		\begin{scope}
			\draw [black, ->] (0,0)-- (1.2,0);
			\draw [black, ->] (0,0)-- (0,1.1);
			\draw [black, ->] (0,0)-- (-.5,-.5);
			\draw (0,0) circle [radius=1];
			\draw[fill=red] (0,0) circle [radius=.05];
			\draw (0,0) circle (1 and .3);
			\node [above] at (0,1) {$\ket{0}\!\bra{0}$};
			\node [below] at (0,-1) {$\ket{1}\!\bra{1}$};
		\end{scope}
	
		\end{tikzpicture}
		\caption{The Bloch ball representation of a qubit: The unit sphere represents the pure states and its interior the mixed states. The fully mixed state is the red dot. Orthogonal states are antipodal.}\label{f:bloch}
		\end{figure}

{The set of states of a single qubit is geometrically a ball, the Bloch ball \cite{nc}:   The density matrix  $\rho$ is $2\times 2$:}
\be\label{e:bb}
\rho(\mathbf{x})=\frac{\id +\mathbf{x}\cdot \ \boldsymbol{\sigma}}2, \quad \mathbf{x}\in\mathbb{R}^3,
\ee
{with   $\boldsymbol{\sigma}=(\sigma_1,\sigma_2,\sigma_3)$,  the vector of $2\times 2$  (Hermitian, traceless) Pauli matrices.  
$\rho\ge 0$, provided $ |\mathbf{x}|\le 1$.   The unit sphere, $|\mathbf{x}|^2=1$, represents pure states where $\rho$ is a rank one projection.
The interior of the ball describes mixed states and the center of the ball  the fully mixed state, (Fig.~\ref{f:bloch}).}

 {The geometry of a qubit is not always a good guide to the geometry of general quantum states: $n$-qubits {\em are not} represented by $n$ Bloch balls}\footnote{$n$ Bloch balls describe uncorrelated qubits.},  {and quantum states are not, in general, a ball in high dimensions.}  

Quantum states are mixtures of pure states.
We denote the set of quantum states in an $N\ge 2$ dimensional Hilbert space {$\cal H$} by $D_N$:
\begin{equation}\label{e:mix-pure}
   D_N=\left\{\rho\,\Bigg |\,\rho= \sum_{j=1}^k p_j \ket{\psi_j}\bra{\psi_j}, \ p_j\ge 0,\ \sum_j p_j=1,\ \ket{\psi_j}\in 
   {\cal H},\ \|\psi_j\|=1 \right\}
\end{equation}

\begin{figure}
    \centering
    \includegraphics[width=.7 \textwidth]{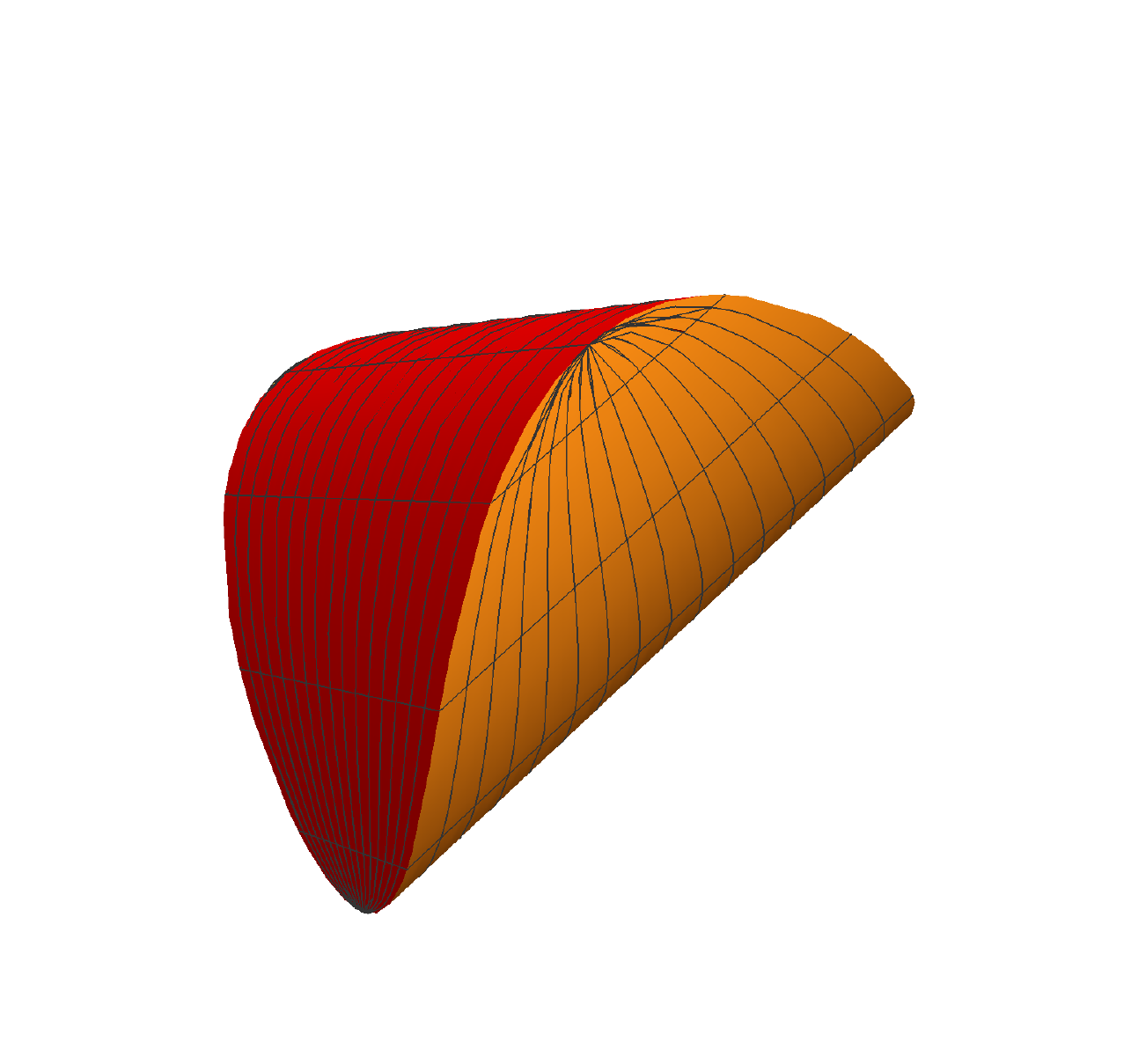}
    \caption{The figure shows a three dimensional convex body whose geometry shares some of the qualitative features of  $D_N$ (and $S_{N_1,\dots,N_k}$): The extreme points lie  on a low dimensional smooth sub-manifold of the boundary at fixed distance from the center and lack inversion symmetry.}
    \label{f:pringles}
\end{figure}

The representation implies:
\begin{itemize}
    \item The quantum states form a convex set.
    \item The pure states are its extreme points. {As we shall see in section \ref{s:bdry}, the set of pure states is a smooth manifold, which is a tiny subset of $\partial D_N$  when $N$ is large.}
    \item The spectral theorem gives (generically) a distinguished decomposition  with $k\le \dim {\cal H}$. 
\end{itemize}
{Fig.~\ref{f:pringles} shows a three dimensional body whose geometry reflects better the geometry of the space of states $D_N$  for general $N$, (and also the spaces of separable states), than the Bloch sphere does.}

Choosing a basis in ${\cal H}$, the state $\rho$ is represented  by a positive $N\times N$ matrix with unit trace ($N=\dim {\cal H}\ge 2$). 
In the case of  $n$-qubits $N=2^n$.  Since the sum of two positive  matrices is a positive matrix, the positive matrices form a convex cone in $\mathbb{R}^{N^2}$, and the positive matrices with unit trace are a slice of this cone. The slice is an $N^2-1$ dimensional convex body with the pure states  $\ket{\psi}\bra{\psi}$ as its extreme points {and the fully mixed state as its ``center of mass"}.

The geometric properties of $D_N$ can be complicated and, because of the high dimensions involved, counter-intuitive. Even the case of two qubits, where $D_N$ is 15 dimensional, is difficult to visualize \cite{zb,myrheim,ak,as}.

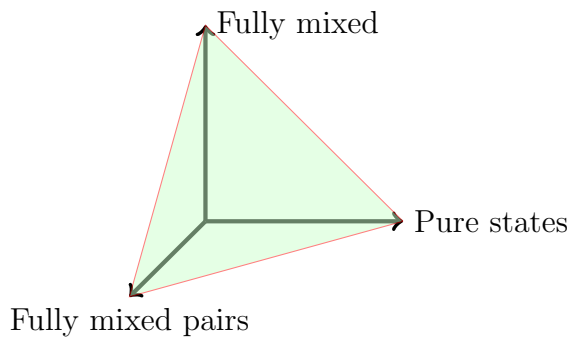
\begin{figure}[ht!]
\centering{}
\begin{tikzpicture}[scale=2]
		\begin{scope}
			\draw [ultra thick, black, ->] (0,0)-- (1.3,0);
			\draw [ultra thick,black, ->] (0,0)-- (0,1.3);
			\draw [ultra thick,black, ->] (0,0)-- (-.5,-.5);
			\draw [fill=green!20!white,draw=red,opacity=.5] (1.3,0) --(0,1.3)--(-.5,-.5) --(1.3,0);
			\node [right] at (0,1.3) {Fully mixed} node [right] at (1.3,0) {Pure states} node [below] at (-.5,-.5) {Fully mixed pairs};
				\end{scope}
	
		\end{tikzpicture}
		\caption{The equivalence classes of qutrits make  a triangle.}\label{f:equiv-state}
		\end{figure}

In contrast with the complicated geometry of $D_N$, the geometry of equivalence classes of quantum states under unitaries, even for large $N$, is simple: It is  parametrized by eigenvalues and represented by the $N-1$-simplex, Fig. \ref{f:equiv-state},
\[
1\ge \rho_1\ge \dots \ge \rho_N\ge 0, \quad \sum\rho_j=1
\]
All pure states are represented by the single extreme point, $(1,0,\dots,0)$,  and the fully mixed state by the extreme point $(1,\dots, 1)/N$. The equivalence classes corresponding to the Bloch ball are represented by an interval (1-simplex) which corresponds to the radius of the Bloch ball.

 {Clearly, the geometry of $D_N$ does not resemble} the geometry of the set of equivalence classes:  {The two live in different dimensions, have different extreme points and  $D_N$ is, of course,  not a polytope.}

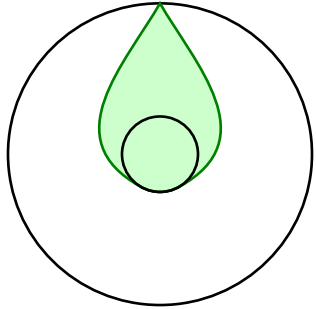
\begin{figure}[ht!]
    \begin{center}
		\begin{tikzpicture}[scale=6/3,line width=1pt,black]	
		\draw (0,0) circle (1);
		\draw (0,0) circle (1/4);
  	    	\draw[fill=green!20!white, draw=green!50!black] (-120:.25) arc (-120:-60:.25) .. controls (.7,.1) and (.23,.6) ..(0,1).. controls (-.23,.6) and (-.7,.1) ..(-120:.25);
  		\draw (0,0) circle (1/4);
	 \end{tikzpicture}
	 \caption{The  inscribed, Gurvits-Barnum, ball is represented by the small circle and the bounding sphere  by the large circle. The green area represents $D_N$.  }\label{f:r2R}
	\end{center}
	\end{figure}

{One of the features of a qubit that holds for any $D_N$,  is that the pure states are equidistant from the   fully mixed state. Indeed  }
\begin{equation}\label{e:pure-mixed-dis}
    Tr \left(\ket\psi\bra\psi-\frac \id N\right)^2= 1-\frac 1N
\end{equation}
{}{This implies that $D_N$ is contained in a ball, centered at the fully mixed state, whose radius squared is given by the rhs of Eq.~(\ref{e:pure-mixed-dis}).}
{$D_N$ is, however, increasingly unlike a ball when $N$ is large:} {}{The} largest ball inscribed in $D_N$ is {the Gurvits-Barnum ball\footnote{Gurvits and Barnum define the radius of the ball by its purity, rather than the distance from the maximally mixed states. } \cite{GB}}:
\begin{equation}\label{e:bgb}
     B_{gb}=\left\{\rho\,\Bigg|\, 
    Tr\left(\rho-\frac\id N\right)^2\le \frac 1 {N-1}-\frac 1 N\right\}
\end{equation}
{It is a tiny ball when $N$ is large, whose radius squared is given by the rhs of Eq.~(\ref{e:bgb}) and centered at the fully mixed state. }
It follows that:
\begin{itemize}
    \item \[
    \dim D_N=N^2-1
    \]
    \item Since
    \be\label{e:ratio}
 \frac{\text{radius of bounding ball of } D_N}{\text{radius of inscribed ball of }D_N }= {N-1}
\ee
only $D_2$ is a ball and $D_N$ gets {increasingly far from a ball}
when $N$ is large, Fig. \ref{f:r2R}.
\end{itemize}

Another significant difference between a single qubit and the general case is that $D_N$ is inversion symmetric only for $N= 2$.  Indeed, inversion with respect to the fully mixed state is defined by 
\begin{equation}
    I(\rho)=\frac{\id}{N}-\left(\rho-\frac{\id}{N}\right)=\frac 2 N \id -\rho
\end{equation}
Evidently $I$  is trace preserving and $I^2=1$. However, it is not positivity preserving  {unless the purity of $\rho$ is sufficiently small. Indeed, $\rho\ge0$ and $I(\rho)\ge 0$ imply} that\footnote{{$I(\rho)\ge 0$ iff $\|\rho\|\le 2/N$}.}
\begin{equation}\label{e:no-inversion}
    0\le Tr \big(\rho I(\rho)\big)=\frac 2 N  -Tr(\rho^2)
\end{equation}

{The low symmetry together with the large aspect ratio indicate that the geometry of $D_N$ may be complicated. It can be visualized, to some extent, by looking at cross sections. As we shall see $D_N$ has several cross sections that are simple to describe:}  Regular simplexes, balls and  hyper-octahedra.

$D_N$ has a  Yin-Yang relation to spheres in high dimensions:  As $N$ gets large $D_N$ gets increasingly far from a ball as  is evidenced by the diverging  ratio of the bounding sphere to the inscribed ball. At the same time there is a sense in which the converse is also true: Viewed from the center (the fully mixed state), the distance to $\partial D_N$, is the same for almost all directions. In this sense, $D_N$ increasingly resembles a ball. {The radius of the ball can be easily computed using standard facts from random matrix theory \cite{mehta}, and we find that as $N\to\infty$, for almost all directions,} 
    \be
2NTr\left(\rho-\frac\id N\right)^2\to 1, \quad \rho\in\partial D_N
    \ee
The fact that $D_N$ is almost a ball is not surprising. In fact, a rather general consequence of the theory of ``concentration of measures", \cite{alon,szarek,milman}, is that sufficiently nice high dimensional convex bodies are essentially balls\footnote{Applications of concentration of measure to quantum information are given in e.g. \cite{GB,szarek,hayden}. }. $D_N$, however,  is not sufficiently nice so that one can simply apply standard theorems from concentration of measure. {Instead, we use information about the second moment of $D_N$ and H\"older inequality to show that the set of directions $\theta$ that allow for states with significant purity has super-exponentially small measure (see section \ref{s:concentration}).}

\subsection{The geometry of separable states}
{The Hilbert space of a quantum system partitioned into $n$ groups of (distinguishable) particles has a tensor product structure ${\cal  H}={\cal H}_{N_1}\otimes\dots\otimes{\cal H}_{N_n}$. The set of separable states of such a  system, denoted $S_{N_1\dots N_n}$,  is defined  by \cite{PBook}}, 
\be\label{e:mix-separable}
S_{N_1\dots N_n}=\left\{\rho\ \Bigg|\,\rho= \sum_{j=1}^k p_j \ket{\psi_j^{(1)}}\bra{\psi_j^{(1)}}\otimes
\dots \otimes\ket{\psi_j^{(n)}}\bra{\psi_j^{(n)}},\,\ket{\psi_j^{(m)}}\in {\cal H}_{N_m}\right\}
\ee
{where $p_j$ are probabilities. }

 {For reasons that we shall explain in section \ref{s:s}}, $S_{N_1\dots N_n}$ are more difficult to analyze than $D_N$. They  have been studied by many authors from different perspectives  \cite{as,szarek,GB,zb}. It will be a task with diminishing returns to try and make a comprehensive list of all of the known results. Selected few references are \cite{myrheim,ak,szarek,zh,hayden,bruss,doherty,hhh,stormer}.
We shall review, instead, few elementary observations and accompany them by pictures. 

 {The representation in Eq.~(\ref{e:mix-separable}) implies  that}
\begin{itemize}
    \item $S_{N_1\dots N_n}\subseteq D_N$ where $N=\prod_j N_j$
    
    \item $S_{N_1\dots N_n}$ is a convex set with pure-product states as its extreme points.
    \item {The  finer the partition the smaller the set}
    \[
    S_{N_1,N_2,N_3}\subseteq S_{N_1,N_2\times N_3}\cap S_{N_3,N_1\times N_2}\cap S_{N_2,N_1\times N_3}
    \]
{Strict inclusion implies that Alice, Bob and Charlie may have a 3-body entanglement that is not visible in any bi-partite partition.} 
    \item  By Caratheodory theorem, one can always find a representation with $k\le N^2$ in Eq.~(\ref{e:mix-separable}). In the case of two qubits a results of Wootters \cite{wootters} gives $k\le N$. 
    \item $S_{N_1\dots N_n}$ {is invariant under partial transposition, (transposition of  {any one} of its factor), i.e.}
    \be
    \ket{\psi_j^{(m)}}\bra{\psi_j^{(m)}}\mapsto
    \left(\ket{\psi_j^{(m)}}\bra{\psi_j^{(m)}}\right)^t
    \ee
\item The bounding sphere of $S_{N_1\dots N_n}$ is the bounding sphere of $D_N$.
 \item {The separable states are of full measure:}
 \be\label{e:dims}
\dim S_{N_1,\dots,N_n}=N^2-1
\ee
{It is enough to show this for the maximally separable set. For simplicity, consider the case of $n$ qubits. For each  qubit ${\id+ \sigma_\mu}$ with $\mu=1,\dots, 3$ are linearly independent and positive. The same is true for their tensor products. This gives $4^n=N^2$ linearly independent separable states spanning a basis in the space of Hermitian matrices.} 
\end{itemize}

{By a result of \cite{GB}:}
 \be\label{e:gb}
 B_{gb}\subseteq S_{N_1, N_2}
 \ee
 for any partition. It implies that:
\begin{itemize}
\item Since\footnote{\label{fn:center}The balls are centered at the mixed state which is the natural ``center of mass" of all states.}
\be
\frac{\text{radius of bounding ball of } S_{N_1, N_2}}{\text{radius of inscribed ball of } S_{N_1,N_2}}=N-1
\ee
\end{itemize}
the separable states get increasingly far from a ball when $N$ is large.


 {We expect that the separable states too are approximated by  balls in most directions, but unlike the case of $D_N$, we do not know how to estimate the radii of these balls. }

\section{Two qubits}
{Two qubits give a much better intuition about the geometry of general quantum states than a single qubit. However, as 2 qubits live 
in 15 dimensions, they are still hard to visualize. 

{One way to gain insight into the geometry of two qubits is to consider equivalence classes that can be visualized in 3 dimensions \cite{hh,zb,ak,mbr}). However, as we have noted above, the geometry of equivalence classes is distinct from the geometry of states. An alternate way to visualize 2 qubits is to look at  2 and 3 dimensional cross sections through the space of states. }

{The states of two qubits can be parametrized by $\mathbf{x}\in \mathbb{R}^{15}$. It is convenient to lable the 15 components of $x$ by  $\mathbf{x}_{\mu\nu}$ with $\mu\nu\in({01},\dots,{33})$ }
\be
\rho(\mathbf{x}) =\frac {\id_4+\sqrt{3}\,\sum x_{\mu\nu}{\sigma}_\mu\otimes \sigma_\nu}4,
\ee
{${\sigma_\mu}$ are the Pauli matrices. By a 2 dimensional section in the space of two qubits we mean a two dimensional plane in $\mathbf{R}^{15}$ going through the origin.} 
\subsection{Numerical sections for 2 qubits}

The 2 dimensional figures \ref{f:2dimRan} and \ref{f:2dimPur} show random sections obtained by numerically testing the positivity and separability of $\rho$, using Mathematica.
{A generic plane will miss the pure states, which are a set of lower dimension. This situation is shown in Fig.~\ref{f:2dimRan}.} 

\begin{figure}[!ht]
  
  \centering
   \includegraphics[width=0.4\textwidth]{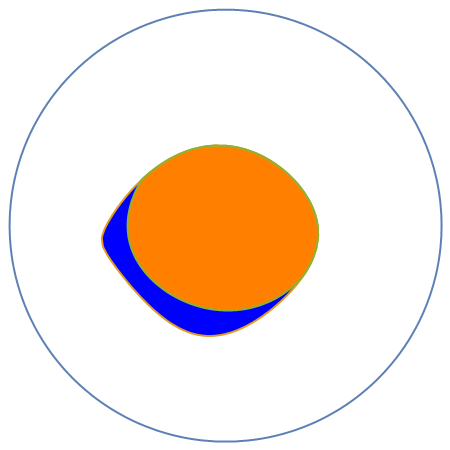} 
   \caption{A numerical computation of a random planar cross section through the origin in the   space of 2 qubits. The orange spheroid shows the separable states and the blue moon the entangled states. The orange spheroid is not too far from a sphere centered at the origin. }\label{f:2dimRan}
\end{figure}

{Fig.~\ref{f:2dimPur} shows a two dimensional section obtained by picking two pure states randomly. Since a generic pure state is entangled,  the section goes through two pure entangled states lying on the unit circle. }

\begin{figure}[!ht]
  \centering
    \includegraphics[width=0.4\textwidth]{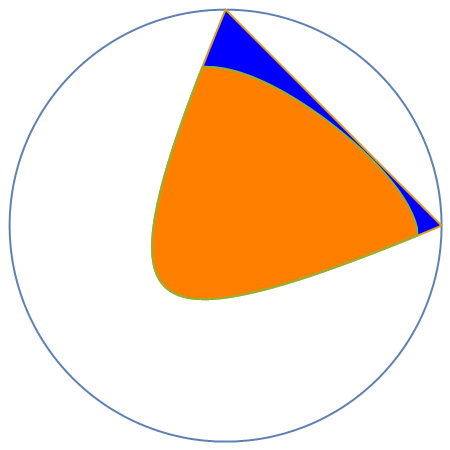}
   \caption{A numerical computation of a planar cross section through the origin and through two random pure states (located on the circle).  The orange region shows the separable states and the blue region the entangled states. The section is quite far from a sphere centered at the origin.  }\label{f:2dimPur}
\end{figure}

\subsection {A 3-D section through Bell states}

Consider the 3D cross section  given by\footnote{A generic two qubits  state is SLOCC equivalent to a point of this section, see \cite{ak}.}:
\be\label{e:rhos}
4\rho(x,y,z)=  {\id}   + \sqrt 3\big( x \sigma_1\otimes \sigma_1+
y \sigma_2\otimes \sigma_2+z \sigma_3\otimes \sigma_3\big)
\ee
The section has the property that both subsystems are maximally mixed
\[
Tr_2\rho=Tr_1\rho=\frac{\id}{2}
\]
Since the purity is given by
\be
 Tr (\rho^2)= \frac 1 4+\frac 3 4\big( x^2+y^2+z^2\big)
\ee
{the pure states lie on the unit sphere and all the states in this section must lie inside the unit ball.}

The matrices on the right commute and satisfy one relation
\[
(\sigma_1\otimes \sigma_1)\,(\sigma_2\otimes \sigma_2)=-\sigma_3\otimes \sigma_3
\] 
It follows that $\rho\ge 0$  iff $(x,y,z)$ lie in the intersection of the 4 half spaces:
\be
 x\pm y\mp z\ge -\frac 1 {\sqrt 3}, \quad  -x\pm y\pm z\ge -\frac 1 {\sqrt 3}
\ee
This defines a regular tetrahedron with vertices
\be
\frac 1 {\sqrt 3} ( 1, 1,-1), \quad \frac 1 {\sqrt 3} ( 1, -1,1)\quad
\frac 1 {\sqrt 3} ( -1, 1,1),\quad \frac 1{\sqrt 3} (-1,-1,- 1)
\ee
{The vertices of the tetrahedron lie at the corners of the cube in Fig. \ref{f:platonic}, at unit distance from the origin. It follows that the vertices of the tetrahedron represent pure states.}
As the section represents states with maximally mixed subsystems, the four pure states are maximally entangled:  They are the 4 Bell  states

The  pairwise averages of the four corners of the tetrahedron give the  $\binom{4}{2}=6 $ vertices of the octahedron in Fig.~(\ref{f:platonic}). By Eq.~(\ref{e:2bell-sep}) below,  {these averages represent separable states. It follows that }
the octahedron represents separable states.
\begin{figure}[!ht]
\begin{center}
\tdplotsetmaincoords{60}{120}
\begin{tikzpicture}[scale=2,
		tdplot_main_coords]
			
	\coordinate (c1) at (-1,-1,-1);	
	\coordinate (c2) at (1,-1,-1);	
	\coordinate (c3) at (1,1,-1);	
	\coordinate (c4) at (-1,1,-1);	
	\coordinate (c5) at (-1,-1,1);	
	\coordinate (c6) at (1,-1,1);	
	\coordinate (c7) at (1,1,1);	
	\coordinate (c8) at (-1,1,1);	
	
	\draw [thick,black] (c1) -- (c2) -- (c3) -- (c4)-- cycle;
	\draw [thick,black] (c5) -- (c6) -- (c7) -- (c8)-- cycle;
	\draw [thick,black] (c1) -- (c5);
	\draw [thick,black] (c2) -- (c6);
	\draw [thick,black] (c3) -- (c7);
	\draw [thick,black] (c4) -- (c8);
	

	\draw[fill=green!20!white,draw=red,opacity=.5] (c1) -- (c3) -- (c6) -- cycle;
	\draw[fill=green!20!white,draw=red,opacity=.5] (c6) -- (c8) -- (c3) -- cycle;
	\draw[fill=green!20!white,draw=red,opacity=.5](c1) -- (c6) -- (c8) -- cycle;
	
	\coordinate (o1) at (0,0,1);	
	\coordinate (o2) at (0,1,0);	
	\coordinate (o3) at (1,0,0);	
	\coordinate (o4) at (0,0,-1);	
	\coordinate (o5) at (0,-1,0);	
	\coordinate (o6) at (-1,0,0);	

	\draw[fill=blue!20!white,draw=red,opacity=.3] (o1) -- (o2) -- (o3) -- cycle;
	\draw[fill=blue!20!white,draw=red,opacity=.3] (o1) -- (o2) -- (o6) -- cycle;
	\draw[fill=blue!20!white,draw=red,opacity=.3] (o1) -- (o5) -- (o6) -- cycle;
	\draw[fill=blue!20!white,draw=red,opacity=.3] (o1) -- (o5) -- (o3) -- cycle;
	\draw[fill=blue!20!white,draw=red,opacity=.3] (o4) -- (o5) -- (o6) -- cycle;
	\draw[fill=blue!20!white,draw=red,opacity=.3] (o4) -- (o5) -- (o3) -- cycle;
	\draw[fill=blue!20!white,draw=red,opacity=.3] (o4) -- (o2) -- (o6) -- cycle;
	\draw[fill=blue!20!white,draw=red,opacity=.3] (o4) -- (o2) -- (o3) -- cycle;
	
\end{tikzpicture}
  \caption{A 3D cross section in the space of states of 2 qubits. The 4 Bell states lie at the vertices of the tetrahedron of states. The octahedron shows the separable states. {The vertices of the cube represent the extreme points of the, trace normalized, entanglement witnesses. The cube is inscribed in the unit sphere (not shown).} }\label{f:platonic}
\end{center}
\end{figure}
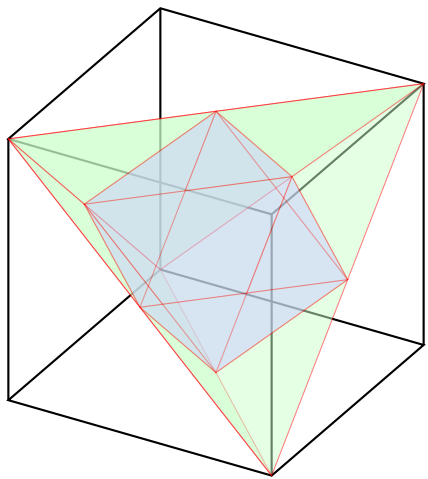

The cube,  represents the trace-normalized entanglement witnesses (see section \ref{s:witness}). A vector $\mathbf{s}$ inside the octahedron {represents the state $\rho(\mathbf{s})$ per Eq.~(\ref{e:rhos}). Similarly,   a vector $\mathbf{w}$ inside the cube represents the witness $W$ (see Eq.~(\ref{e:witness-coord})). Since the cube is the dual of the octahedron we have}
\be
0\le Tr \big(W\rho(\mathbf{s})\big)=\frac { 1+ 3 \mathbf{s\cdot w}} 4\,,
\ee
{which is the defining relation of witnesses.}

\section{Basic geometry of Quantum states}
		\subsection{Choosing coordinates}\label{s:coordinatees}
  {Any Hermitian $N\times N$ matrix with unit trace can be written as: }
\be\label{e:sigma-coord}
\rho(\mathbf{x}) =\frac {\id_N+\sqrt{N-1}\,  \mathbf{x}\cdot \boldsymbol{\sigma}}N,\quad \mathbf{x}\in\mathbb{R}^{N^2-1}
\ee
$\id_N=\sigma_0$ is the identity matrix and  
 $\boldsymbol{\sigma}=(\sigma_1,\dots,\sigma_{N^2-1})$
a vector of $N^2-1$ traceless, Hermitian,  mutually orthogonal, $N\times N$   matrices 
\be\label{e:orthogona}
Tr\, ({\sigma}_\alpha{\sigma}_\beta)=\delta_{\alpha \beta}N, \quad \alpha,\beta\in \{1,\dots, N^2-1\}
\ee
{This still leaves considerable freedom in choosing the coordinates  $\sigma_\alpha$ and one may impose additional desiderata. For example:}
\begin{itemize}
\item   {$\sigma_\alpha$ are either real symmetric or imaginary anti-symmetric }
\be\label{e:symm-anti}
\sigma_\alpha^t=\pm \sigma_\alpha
\ee
This requirement is motivated by $\rho\ge 0 \Longleftrightarrow \rho^t\ge 0$
\item   $\sigma_\alpha $ for $\alpha\neq 0$ are unitarily equivalent, i.e. are iso-spectral. 
\end{itemize}
A coordinate system that has these properties in $N=2^n$ dimensions, is the  {(generalized)} Pauli coordinates: 
\be\label{e:sigma_mu}
{\sigma}_{\mu}= \sigma_{\mu_1}\otimes\dots\otimes\sigma_{\mu_n}, \quad \mu_j\in\{0,\dots,3\}, \quad \mu\in\{1,\dots,N^2-1\}
\ee
 {$\sigma_\mu$ are iso-spectral with eigenvalues $\pm 1$. This follows from}:
\be
{\sigma}_\mu^2=\id_N, \quad Tr \sigma_\mu= 0
\ee
 {The Pauli coordinates behave nicely under transposition:}
\be
\sigma_\mu^t=\pm \sigma_\mu
\ee
In addition, they either commute or anti-commute 
\be \label{e:c} 
{\sigma}_\mu {\sigma}_\beta=\pm {\sigma}_\beta {\sigma}_\mu\,.
\ee
 This will prove handy in what follows. One drawback of the Pauli coordinates is that they only apply to  Hilbert spaces with special dimensions, namely $N=2^n$.
 
For $N$  {arbitrary, one may not be able satisfy all the desiderata simultaneously.} {In particular}, the standard basis
\be
X_{jk}=\ket{j}\bra{k}+ \ket{k}\bra{j} ,\quad Y_{jk}= i(\ket{j}\bra{k}- \ket{k}\bra{j}),\quad
Z_{jN}=\ket{j}\bra{j}- \ket{N}\bra{N} 
\ee    
is iso-spectral with eigenvalues $\{\pm 1,0\}$ and behave nicely under transposition.     {However, the $Z_{jN}$ coordinates are not mutually orthogonal}. 

{With a slight abuse of notation we redefine}
\be
D_N=\left\{\mathbf{x}\Big|\rho(\mathbf{x})\ge 0\right\}
\ee

Note that the Hilbert space and the Euclidean distances are related by scaling 
\be\label{e:hilbert-euclid}
N Tr (\rho-\rho')^2=(N-1) \left(\mathbf{x}-\mathbf{x}'\right)^2
\ee

The basic geometric properties of $D_N$ follow from Eq.~(\ref{e:mix-pure}): 
\begin{itemize}
    \item The fully mixed state, $\id/N$, is represented by the origin $\mathbf{x}=0$ 
\item The pure states lie on the unit sphere for all $N$. This follows {either from Eqs.(\ref{e:pure-mixed-dis},\ref{e:hilbert-euclid}) or, alternatively, from a direct computation of the purity}: 
\be\label{e:purity}
p=Tr \,(\rho^2)=\frac 1 N +\left(1-\frac 1N\right) |\mathbf{x}|^2
\ee

\item   Since the pure states are the extreme points of $D_N$:
\be
D_N\subseteq B_1= \left\{\mathbf{x}\Big||\mathbf{x}|\le 1 \right\}
\ee
    \item Since $\rho\ge0 \Longleftrightarrow \rho^t\ge 0$ $D_N$ is  symmetric under reflection of the ``odd'' coordinates associated with the anti-symmetric matrices. 
    \item Since there is no reflection symmetry for the ``even'' Pauli coordinates, $\sigma_\alpha=\sigma_\alpha^t$, one does not expect $D_N$ to have inversion symmetry in general (as we have seen in Eq.~(\ref{e:no-inversion})).  
     
\end{itemize} 

{Let $\theta$ be a point on the unit sphere in $\mathbb{R}^{N^2-1}$ and $(r,\theta)$ be the polar representation of $\mathbf{x}$, in particular $r=|\mathbf{x}|$. Denote by $r(\theta)$ the radius function of $D_N$, i.e. the distance from  the origin of the boundary of $D_N$ in the $\theta$ direction. Then}
\begin{equation}\label{e:radius-fn}
   1\ge r(\theta)=-\frac{r_0}{ \lambda_1(\theta)}>0, \quad 1\ge r_0=\frac 1{\sqrt{N-1}}\ge 0
\end{equation}
where $\lambda_1(\theta)<0$ is the smallest eigenvalue of \begin{equation}\label{e:S}
 S(\theta)= \boldsymbol{\theta}\cdot \boldsymbol{\sigma}  
\end{equation}


\subsection{Most of the unit sphere does not represent states}\label{s:bdry}

For $N=2$, every point of the  unit sphere represents a pure state, however, for $N\ge 3$ this is far from being  the case. In fact, $\rho(\mathbf{x})$ of Eq.~(\ref{e:sigma-coord}) is not a positive matrix for most $\mathbf{x}^2= 1$. This follows from a simple counting argument: 
Pure states can be written as  $\ket{\psi}\!\bra\psi$ with $\ket\psi$ a normalized vector in $ \mathbb{C}^N$.  It follows that
\be
\dim_{\mathbb{R}} (\text{pure states})=2(N-1)
\ee
while
\be
\dim \left(\text{unit sphere in }\mathbb{R}^{N^2-1}\right)=N^2-2
\ee
When $N\ge 3$  pure states make a small subset of the  of the unit sphere. When $N$ is large the ratio of dimensions   is arbitrarily small.

Since (pure) states make a tiny subset of the unit sphere,  spheres with radii close to $1$, should be mostly empty of states. In section \ref{s:concentration} we shall give a quantitative estimate of this observation.    

\subsection{Inversion asymmetry}

The Hilbert space and the Euclidean space scalar products are related by 
\be\label{e:hs}
N\,Tr\,(\rho\rho')=1+(N-1) (\mbf{x\cdot x}')
\ee 
The positivity  of  $Tr \rho\rho'\ge 0$ and Eq.~(\ref{e:hs}) say that if both $\mathbf{ x}$ and $\mathbf{ x'}$ correspond to bona-fide states  then it must be that 
\be\label{e:hs3}
\mathbf{x\cdot x'}\ge - \frac 1 {N-1}, \quad \mathbf{x,x'}\in D_N
\ee
In particular, no two pure states are ever related by inversion if $N\ge 3$, (see fig. \ref{f:tau}). 

\begin{figure}[ht!]
\begin{center}
			\begin{tikzpicture}[scale=6/3,line width=1pt,black]	
				  \filldraw[fill=green!20!white, draw=green!50!black]	
							 (190:1) arc (190:-10:1);
  				\draw (0,0) circle (1 cm);
  				\node at (0,1.1) {$\ket{\psi}\!\bra{\psi}$};
  				\node at (0,-.4) {$\tau_\psi$};
  				\draw[fill,red] (0,0) circle (.05 cm);
  					\draw[fill,blue] (0,-.16) circle (.03 cm);
			 \end{tikzpicture}
			 \caption{The figure shows the intersection of the half space in  Eq.~(\ref{e:hs3}) with the unit ball in the  case that $\mathbf{ x}$ corresponds to a pure state $\ket\psi$.    {The blue dot represents the state $\tau_\psi$ of Eq.~(\ref{e:tau})}.}\label{f:tau}
			 \end{center}
	  \end{figure}
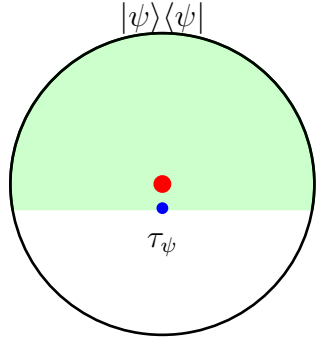

\subsection{The inscribed sphere}\label{s:inscribed}

{The inscribed ball in $D_N$,  the Gurvits-Barnum ball, is } 
\begin{equation}\label{e:gb-ball}
B_{gb}= \left\{\mathbf{x}\Bigg||\mathbf{x}|\le\frac 1 {N-1} \right\}\subseteq D_N
\end{equation}
It is easy to see that the inscribed ball is at most the Gurvits-Barnum ball since the state
\be\label{e:tau}
\tau_\psi=\frac{\id- \ket\psi\bra\psi}{N-1}
\ee
clearly lies on $\partial D_N$. {Using Eq.~(\ref{e:hs}), one verifies that $|\mathbf{x}_\psi|=\frac 1 {N-1}$ saturating Eq.~(\ref{e:gb-ball}).}

    {Eq.~(\ref{e:gb-ball}) follows from:}
\begin{align}
 B_{gb}&= \left\{\mathbf{x}\Bigg| \mathbf{x}\cdot \mathbf{x'} \ge - \frac 1 {N-1},\mathbf{|x'|}\le 1 \right\}\nonumber \\
 &\subseteq  \left\{\mathbf{x}\Bigg| \mathbf{x}\cdot \mathbf{x'} \ge - \frac 1 {N-1},\mathbf{x'}\in D_N  \right\}\nonumber
 \\
 &{\subseteq}D_N
\end{align}
    {In the last step we used the fact that the positivity of $\rho$ follows from the positivity of $\rho'$ by  Eq.~(\ref{e:hs}).}

\begin{rem}
An alternate proof is:
{By Eq.~(\ref{e:radius-fn}) minimizing $r(\theta)$ is like minimizing the smallest eigenvalue $\lambda_1(\theta)$ of $S(\theta)=\boldsymbol{\theta}\cdot\boldsymbol{\sigma}$ under the constraints  $Tr S^2(\theta)=N$ and $Tr S(\theta)=0$. The minimum occurs when} 
\[ spec(S)=\left(-\frac 1{r_0}, r_0, \dots,r_0\right),\quad r_0=\frac 1 {\sqrt{N-1}}
\]  
This, together with Eq.~(\ref{e:radius-fn}), gives $r_0^2$ for the radius of the inscribed  ball of non-negative matrices.
\end{rem}

\section{Cross sections}
$D_N$ has few sections that are simple to describe, even when $N$ is large.
\subsection{Cross sections that are N-1 simplexes}\label{s:ortho}

Let $\mathbf{v}_j$, with $j=0,\dots, N-1$ be the (unit) vectors associated with pure states $\rho_j=\ket{\psi_j}\!\bra{\psi_j}$ corresponding to the orthonormal basis $\{\ket{\psi_j}\}$. Using Eq.~(\ref{e:hs})
 we find for $Tr \rho_j\rho_k=\delta_{jk}$ 
\be\label{e:sp}
\mathbf{v}_j \cdot \mathbf{v}_k=\frac{N\delta_{jk}-1}{N-1}
\ee
For a single qubit, $N=2$, orthogonal states are (annoyingly) represented by antipodal points on the Bloch sphere.  The situation improves when $N$ gets large: Orthogonal states are represented by almost orthogonal vectors.   Moreover, from Eq.~(\ref{e:sigma-coord})
\begin{equation}\label{e:cms}
  \sum_{j=0}^{N-1} \rho_j=\id \Longleftrightarrow 
\sum_{j=0}^{N-1} \mathbf{v}_j=0
\end{equation}
The $N$ vectors  $\mathbf{v}_j$ define a regular $(N-1)$-simplex, centered at the origin (in $\mathbb{R}^{N^2-1}$, see Fig. \ref{f:frame})}
\be
C_{N-1}=(\mathbf{v}_0,\dots,\mathbf{v}_{N-1})
\ee
Since the boundary of $C_{N-1}$ represent states that are not full rank, it belongs to the boundary of $D_N$ and therefore is an $N-1$ slice of $D_N$.
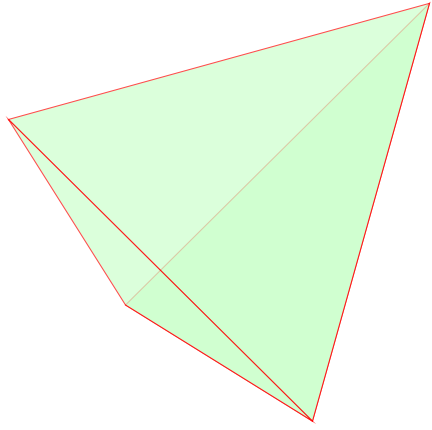
\begin{figure}
\begin{center}

\begin{tikzpicture}[scale=2]
   \begin{scope}[canvas is zx plane at y=1.5]
   \end{scope}
    \begin{scope}[canvas is zx plane at y=-1.5]
   \end{scope}
   \begin{scope}[canvas is zx plane at y=0.0]
   \end{scope}

	\coordinate (c1) at (-1,-1,-1);	
	\coordinate (c2) at (1,-1,-1);	
	\coordinate (c3) at (1,1,-1);	
	\coordinate (c4) at (-1,1,-1);	
	\coordinate (c5) at (-1,-1,1);	
	\coordinate (c6) at (1,-1,1);	
	\coordinate (c7) at (1,1,1);	
	\coordinate (c8) at (-1,1,1);

	\draw[fill=green!20!white,draw=red,opacity=.7] (c1) -- (c3) -- (c6) -- cycle;
	\draw[fill=green!20!white,draw=red,opacity=.7] (c6) -- (c8) -- (c3) -- cycle;
	\draw[fill=green!20!white,draw=red,opacity=.7](c1) -- (c6) -- (c8) -- cycle;

\end{tikzpicture}

  \caption{Any basis of pure states for 2 qubits and in particular, the Bell states and the computational basis are represented by the vertices of a three dimensional tetrahedron. These simplexes are three dimensional cross sections of $D_2$.} \label{f:frame}
\end{center}
\end{figure}
\newpage
\subsection{Cross  sections that are balls}\label{s:clifford}

Suppose $ N=2^{n}$.    {Consider the largest set of  mutually anti-commuting matrices among the $N^2-1$ (generalized) Pauli matrices  ${\sigma}_\alpha$.  Since the Pauli matrices include the matrices  that span a basis of a Clifford algebra  we have at least $\ell$ anti-commuting matrices} 
\be\label{e:clifford}
\{\sigma_j,\sigma_k\}=2\delta_{jk}, \quad j,k\in\{1,\dots,\ell \},\quad\ell=2n+1
\ee
For the anti-commuting $\sigma_j$ we have 
\be
\left(\sum x_j \sigma_j\right)^2=r^2=\sum_j x_j^2 \,
\ee
The positivity of 
\be\label{e:sphere}
\id_N+ \sqrt{N-1}\sum_{j=1}^\ell\,{x}_j\sigma_j\ge 0
\ee 
holds iff
\[
r\le r_0=\frac 1 {\sqrt{N-1}}
\]
This means that $D_N$ has $\ell$ dimensional cross sections that are perfect balls\footnote{This and section \ref{s:tinyBalls} are reminiscent of Dvoretzki-Milmann theorem \cite{milman}.}. 
{This result extends to $2^n\le N<2^{n+1}$.} 
\subsection{Cross sections that are polyhedra and hyper-octahedra}\label{s:hoct}
Consider the set of commuting   matrices $\sigma_\alpha$.  Since these $N-1$ matrices can be simultaneously diagonalized,  the positivity condition on the cross-section 
\[
\id+\sqrt{N-1}\sum_{\alpha=1}^{N-1} x_\alpha \sigma_\alpha\ge 0
\]
{reduces to a set of linear inequalities for $x_\alpha$. {This defines a polyhedron.}

In the case of $n$ qubits,   a set of $n$ commuting  $\sigma_\alpha$ matrices  with no relations is:}
\[
\sigma_\alpha= \id\otimes\dots\id \otimes\sigma_x\otimes \id\dots \otimes \id,\quad \alpha=1,\dots, n 
\]
The cross section is the intersection of $N$ half-spaces
\[
\pi_\alpha\cdot x\ge -r_0
,\quad \alpha=1,\dots,N, \quad \pi_\alpha\in\{-1,1\}^n
\]
The {corresponding cross section} is a regular $n$ dimensional hyper-octahedron\footnote{See footnote \ref{fn:octa}.}:      {A regular, convex polytope with $n$ vertices and $N=2^n$ hyper-planes. (The dual of the $n$ dimensional cube.)
\subsection{2D cross sections in the Pauli basis}

Any two dimensional cross section along {two} Pauli coordinates  can be written as:
\be\label{e:N-bloch}
N\rho(x,y) = \id_N+ {\sqrt {N-1}}(x{\sigma}_{\alpha}+y  {\sigma}_{\beta})
\ee
By Eq.~(\ref{e:c}), $ {\sigma}_{\alpha,\beta}$ either commute or anti-commute.  The case that  they anti-commute is a special case of the Clifford ball of section \ref{s:clifford} where positivity implies
\[
x^2+y^2\le r_0^2=\frac 1  {N-1}
\]
{The case that  $ {\sigma}_{\alpha,\beta}$ commute is a special case of section \ref{s:hoct} where positivity holds if }
\be\label{e:square}
|x|+ |y|<r_0
\ee
{Both are balls, albeit in different metrics,  ($\ell^2$ and $\ell^1$), see Fig.~\ref{f:square-circ}.}
\begin{figure}[ht!]
\centering{} 
\begin{tikzpicture}[scale=.5,line width=1pt,black]
 \usetikzlibrary{calc}
   \draw  [fill=green!20!white,draw=red] (-1,0)--(0,-1)--(1,0)--(0,1)--(-1,0); 
    \draw  [red] circle (4 cm); 
   \draw  [fill=green!20!white,draw=red]   (12,0) circle (1 cm);
   \draw [<->,red] (11,-1.2)--(13,-1.2);
   \node [below] at (12,-1.3) {$2r_0$};
       \draw  [red] (12,0) circle (4 cm); 
   \draw [<->,red] (-1,-1.2)--(1,-1.2);
   \node [below] at (0,-1.3) {$2r_0$};
  \end{tikzpicture} 
  \caption{A two dimensional cross sections of the space of states of $n$ qubits, ${D}_n$,  along the  Pauli coordinates, $ ({\sigma}_\alpha,{\sigma}_\beta)$, is either a tiny square or a tiny disk both of diameter $2r_0$.}\label{f:square-circ}
\end{figure}
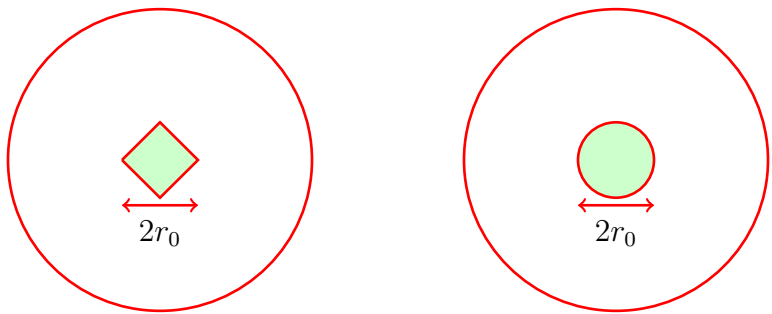
  \section{The radius function}\label{s:radiusFunction}
{By a general principle:  ``All convex bodies in high dimensions are a bit like Euclidean balls" \cite{ball}. More precisely, consider  a convex body $C_N$ in $N$ dimensions, which contains the origin as an interior point.  The radius function of $C_N$ is called  K-Lifshitz, if} 
\be
|r(\theta)-r(\theta')|\le K\|\theta-\theta'\|
\ee
By {a fundamental} result in the theory of concentration of measure \cite{ball,alon},  the radius is concentrated near its median, with a variance that is at most $O(K^2/N)$ \cite{ball}. 

{$D_N$ is a convex body in $N^2-1$ dimensions. As we shall see in the next section, $r(\theta)$ turns out to be $N$-Lifshitz. As a consequence, the variance of the distribution of the radius about the mean is only guaranteed to be $O(1)$. This is not strong enough to conclude that $D_N$ is almost a ball.}


\subsection{The radius function is N-Lifshitz}
Since $D_N$ is convex and {$r(\theta)>0$, the radius function} is   continuous, but not necessarily differentiable. The fact that $D_N$ is badly approximated by a ball is reflected in the continuity properties of $r(\theta)$.

Using the notation of section \ref{s:ortho}, let $C_j$ be the simplex
\begin{equation}
C_{j}=\{\mathbf{v}_0,\dots,\mathbf{v}_{j}\}
\end{equation}
$C_{j}$, for $j<N-1$, is a face of $C_{N-1}$. Denote by $\bar C_{j}$ the bari-center of $C_j$,
\begin{equation}
\bar C_{j}=\frac 1 {j+1}\sum_{\alpha=0}^j \mathbf{v}_\alpha,
\end{equation}
$\bar C_{N-1}=0$ represents the fully mixed state, by Eq.~(\ref{e:cms}).

The three points $\bar C_0, \bar C_{N-2}$ and $\bar C_{N-1}$ define a triangle, shown in Fig.~\ref{f:tr}.        {The sides of the triangle can be easily computed, e.g.} 
\begin{equation}
\bar C_{0}-\bar C_{N-1}=\bar C_{0}
=\mathbf{v}_0 \Longrightarrow |\bar C_{0}-\bar C_{N-1}|= 1
\end{equation}
Since $\mathbf{v}_0$ represents a pure state. Similarly
\begin{equation}
\bar C_{N-2}-\bar C_{N-1}=\bar C_{N-2}=-\frac{1}{N-1}\mathbf{v}_{N-1}
\Longrightarrow |\bar C_{N-2}-\bar C_{N-1}|=\frac{1}{N-1} 
\end{equation}
Consider the path from $\bar C_{N-2}$ to $\bar C_0$. The path lies on the boundary of $D_N$. Therefore $r(\theta)$ in the figure is the radius function. 
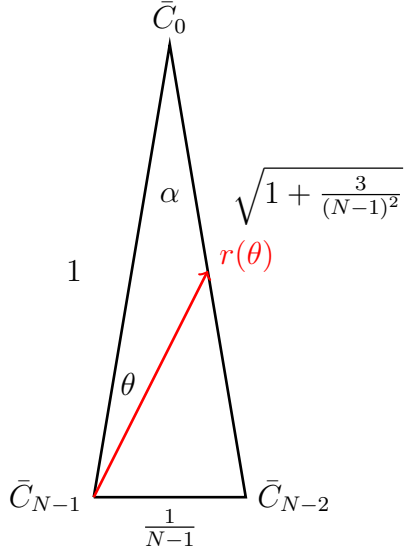
\begin{figure}
    \centering
   			\begin{tikzpicture}[scale=2,line width=1pt,black]	
  				\draw [black] (-1/2,0) --(1/2,0)--(0,3) -- (-1/2,0) node [left] at (-1/2,0) {$\bar C_{N-1}$};
  					\draw [red,->] (-1/2,0) --(1/4,1.5) node [right ] at (1/4,1.6) {$r(\theta)$}
  					node [right,black ] at (1/3,2) {$\sqrt{1+\frac 3 {(N-1)^2}}$};
  				\node at (110:.8) {$\theta$};
  				\node at (90:2) {$\alpha$};
  					\node [above] at (0,3) {$\bar C_0$};
  					\node [right] at (1/2,0) {$\bar C_{N-2}$};
  						\node [left] at (-1/2,1.5) {$1$};
  							\node [below] at (0,0) {$\frac 1 {N-1}$};
			 \end{tikzpicture}
    \caption{ The radius $r(\theta)$ along the path from from $\bar C_{N-2}$ to $\bar C_0$ has large derivative, $O(N)$, near $\theta=0$. }
    \label{f:tr}
\end{figure}
By the law of sines
\begin{equation}
r(\theta)= \frac{\sin\alpha}{\sin(\alpha+\theta)}
\end{equation}
{and}
\begin{equation}
 \left|\frac{r'(\theta)}{r(\theta)}\right|=|\cot (\alpha+\theta)|\Longrightarrow\left|{r'(0)}\right|=  \cot\alpha=\frac{N^2-2N+2}{\sqrt{N(N-2)}}\approx N
\end{equation}
It follows that when $N$ is large, the radius function has large derivatives near the  vertices of the simplex. 
This reflects the fact that {locally} $D_N$ is not well approximated by a ball. 
\begin{rem}
      One can show that $Max\, |r'(\theta)|\le O(N)$ is tight. But we shall not pause to give the proof here.
\end{rem}

\section{A tiny ball in  most directions}\label{s:tinyBalls}
A basic principle in probability theory asserts that while anything that might happen will happen as the system gets large, certain features  become regular, universal,  and non-random \cite{deift,alon}.
{We shall use basic results from random matrix theory \cite{mehta,tw} to show that when $N$ is large $D_N$ approaches a ball whose radius is}
\be\label{e:typical-rad}
r_t\approx \frac {r_0} {2}=\frac {1} {2\sqrt{N-1}}
\ee
Although the radius of the ball is small when $N$ is large, it is much  larger than the inscribed ball {whose radius is $r_0^2$}.

\subsection{Application of random matrix theory}\label{s:mb}

Define a random direction $\boldsymbol{\theta}$  by a vector of iid Gaussian random variables:
\be
\boldsymbol{\theta}=( \theta_1,\dots,\theta_{N^2-1}) ,\quad  \theta_\alpha \sim {\mathcal N}\left[0,\frac 1{N^2-1}\right]
\ee
where ${\mathcal N}\left[\mu,\sigma^2\right]$ denotes the normal distribution with mean $\mu$ and variance $\sigma^2$. \newline
 $\boldsymbol{\theta}$ has mean unit length 
\[
\average{|\boldsymbol{\theta}|^2}=(N^2-1)\average{\theta^2_1}=1
\] 
and small variance 
\begin{align*}
\average{\left(|\boldsymbol{\theta}|^2-1\right)^2}
&{=}-1+\sum_{\alpha\beta} \average{\theta_\alpha^2{\theta}_\beta^2}
\\ &=\frac 2 {N^2-1}
\end{align*}
$S(\boldsymbol{\theta})=\boldsymbol{\theta}\cdot\boldsymbol{\sigma}$ may be viewed as an element of the ensemble of the $N\times N$  traceless Hermitian random  matrices.

By Wigner semi-circle law, when $N$ is large,  the density $\mu$  of eigenvalues  approach a semi-circle
\begin{equation}\label{e:mu}
 d\mu=\frac 2 {\pi \Lambda^2}\sqrt{\Lambda^2-\lambda^2}\, d\lambda
\end{equation}
with edges at 
\be
\Lambda^2= \frac 4 N\average{Tr \,\left(S^2(\boldsymbol{\theta}\right)}=4\ee
{When $N$ is large, the bottom of the spectrum, $\lambda_1(\boldsymbol{\theta})$, a random variable,  is close to the bottom edge at  $-2$  \cite{tw}. 
The radius function $r(\theta)$ is related to the lowest eigenvalue $\lambda_1(\boldsymbol{\theta})$ by Eq.~(\ref{e:radius-fn}) and the value for $r_t$ in Eq.~(\ref{e:typical-rad}) follows}\footnote{Since $\boldsymbol{\theta}$ is a unit vector only on average, there is a slight relative ambiguity in $r$. The error is smaller than the $N^{-2/3}$ fluctuation in the Tracy-Widom distribution \cite{tw}.}.

\begin{rem}
 To see where the value of $\Lambda$ comes from, observe that by Eq.~(\ref{e:mu})
\be
\average{Tr \,\left(\boldsymbol{\theta} \cdot\boldsymbol{\sigma}\right)^2}=\sum_j\average{\lambda^2_j}\approx N\int  \lambda^2 d\mu= N\frac{\Lambda^2} 4
\ee
\end{rem}
\begin{figure}
    \centering
   			\begin{tikzpicture}[scale=6/3,line width=1pt,black]	
  				\draw (0,0) circle (1 cm);
  				\draw[fill=green!20!white, draw=green!50!black] (180:.2) arc (180:360:.2)--(0,1)--(180:.2);
  				\draw[fill,red] (0,0) circle (.05 cm);
			 \end{tikzpicture}
    \caption{$D$ has, in most directions, a small radius, and extends to the pure states only in  rare directions.}
    \label{f:shpitz}
\end{figure}
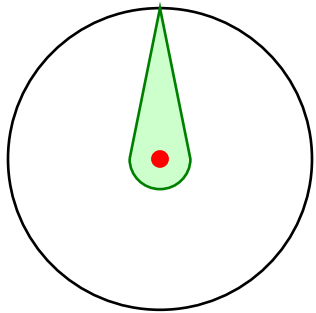
\subsection{Directions associated with states with substantial purity are rare}\label{s:concentration}

{Our aim in this section is to show that the probability for finding directions where  $r(\theta)\ge r_0$ is exponentially small,}\footnote{Note that $r_0=2r_t$ with $r_t$ the radius of the ball determined by random matrix theory. This is an artefact of the method we use where the radius of inertia $r_e$ plays a role. When $N$ is large $r_0\approx r_e$. The stronger result should have $r_0$ replaced by $r_t$ in Eq.~(\ref{e:cf}). } (see Fig. \ref{f:shpitz}). More precisely:} 
\be\label{e:cf}
\text{Prob}(r\ge a)\le\left(\frac {r_0}{a}\right)^{N^2+1}
\ee
In particular, {states that lie outside  the sphere of radius $r_0$} have super-exponentially small measure in the space of directions.

{To see this, let} $d\mu$ be {any} (normalized) measure on $D_N$. 
From Eq.~(\ref{e:purity}) we get a relations between the average purity and the average radius:
\begin{equation}\label{e:okz}
    \int_{D_N}Tr(\rho^2)\, d\mu=\frac 1 N+ \left(1-\frac 1 N\right)\int_{D_N} r^2 d\mu
\end{equation}
In the special case that $d\mu$ is proportional to the Euclidean measure in $\mathbb{R}^{N^2-1}$, {i.e.}
\be
d\mu= \frac 1 {|D_N|}  r^{N^2-2}d\Omega dr
\ee
{where $d\Omega$ is the angular part,}
the lhs {of Eq.~(\ref{e:okz})} 
is known exactly\footnote{In Appendix  \ref{a:mom} we show how to explicitly compute the average purity for measures obtained by partial tracing. } \cite{zs}:
\begin{equation}\label{e:zs}
    \frac 1 {|D_N|}\int_{D_N}Tr(\rho^2)\, dx_1\dots dx_{N^2-1}=\frac {2N}{N^2+1}
\end{equation}
This gives for the {\em radius of inertia} of $D_N$:
\be\label{e:2m}
r_e^2=\int_{D_N} r^2 d\mu= \frac {N+1}{N^2+1}
\ee
We  use this result to estimate the probability of  rare  direction that accommodate states with substantial purity. From Eq.~(\ref{e:2m}) we have
\be
\frac {N+1}{N^2+1}= \frac{\int_{D_N} d\Omega\, dr\,r^2  r^{N^2-2}}{\int_{D_N} d\Omega\, dr r^{N^2-2}}
= \frac{(N^2-1)\int_{D_N} d\Omega\\  r^{N^2+1}(\boldsymbol{\theta})}{(N^2+1)\int_{D_N} d\Omega\ \,r^{N^2-1}(\boldsymbol{\theta})}
\ee
Cancelling common terms we find
\be
\frac {1}{N-1}
= \frac{\int_{D_N} d\Omega\\  r^{N^2+1}(\boldsymbol{\theta})}{\int_{D_N} d\Omega\ \,r^{N^2-1}(\boldsymbol{\theta})}
\ee
By H{\"o}lder inequality  
\be
\int_{D_N} d\Omega \,r^{N^2-1}\le \left(\int_{D_N} d\Omega \,r^{(N^2-1)p}\right)^{1/p}
\left(\int_{D_N} d\Omega \,1^q\right)^{1/q}
\ee
Picking
\[
 p=\frac{N^2+1}{N^2-1}, \quad q=\frac {N^2+1} 2\]
gives
\be 
\int d\Omega \,r^{N^2-1}\le
\left(\int d\Omega \,r^{N^2+1}\right)^{1/p}
\left(\int d\Omega \right)^{1/q}
\ee
And hence,
\begin{align*}
r_0^2=\frac {1}{N-1}&= \frac{\int d\Omega\, r^{N^2+1}}{\int d\Omega \,r^{N^2-1}}\ge
\left(\frac{\int d\Omega\, r^{N^2+1}}{\int d\Omega }\right)^{1/q}
\end{align*}
It follows that
\begin{align*}
r_0^{{N^2+1}}\ge
\frac{\int d\Omega\, r^{N^2+1}}{\int d\Omega }\ge a^{N^2+1} \text{Prob}(r\ge a)
\end{align*}
{This gives Eq.~(\ref{e:cf}).}

\begin{rem}
      The inequality Eq.~(\ref{e:cf}) implies that
\be\label{e:rm}
Prob\left(r^2> r_0^2+\frac a{N^3}\right)\le e^{-a/2}
\ee
independent of random matrix theory. {This implies that $r_t\le  r_0$, which is weaker by factor 2 compared with  Eq.~(\ref{e:typical-rad})). }
\end{rem}


\section{Separable and entangled states}\label{s:s}
\subsection{Why separability is hard}

{Testing whether $\rho$ is a state involves testing the positivity of its eigenvalues. The cost of this computation is polynomial in $N$.  Testing whether $\rho$ is separable is harder.  Properly formulated, it is known to be NP-hard, see e.g. the review \cite{ioannou}.  Algorithms that attempt to decide whether $\rho$ is separable or not have long running times. }

\begin{figure}[ht!]
\centering{} 
\begin{tikzpicture}[scale=2,line width=1pt,black]
\foreach \y in {0,50,130,180,230,310}
   \draw  [draw=red] (120+\y: 1)--(0+\y:1);
    \end{tikzpicture} 
  \caption{A hexagon defined by the intersection of six half-planes.}\label{f:conv}
\end{figure}
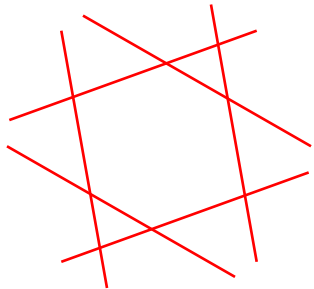

{A pedestrian way to see why separability might be  a hard decision problem is to consider the toy problem of deciding whether a given point $\mathbf{x}\in \mathbb{R}^d$ lies inside a polygon. The polygon is assumed to contain the origin  and is given as the intersection of $M$ half-spaces, each of which contains the origin (see fig. \ref{f:conv}). This can be formulated as}\footnote{$\mathbf{c}_\alpha$ is, in general, not normalized to 1.}
\[
\mathbf{c}_\alpha\cdot \mathbf{x}\le 1,\quad\alpha=1,\dots M
\]
To decide if a point $\mathbf{x'}$ belongs to the polygon, one needs to test $M$ inequalities. The point is that $M$ can be very large even if $d$ is not. For example, in the poly-octahedron $M=2^d$ {and the number of inequalities one needs to check is exponentially large in $d$.}

{Locating a point in a high-dimensional polygon is related to testing for separability \cite{bruss}: The separable states can be approximated by a polyhedron in $\mathbb{R}^{N^2-1}$ whose vertices are chosen from a sufficiently fine mesh of pure product states. Since the number $M$ of half-spaces could, in the general case, be exponentially large in $N$. Testing for separability becomes hard. }

{Myrheim et. al.  \cite{myrheim} gave a  probabilistic algorithm that, when successful,  represents {the input state} as  a convex combination of product states, and otherwise gives the distance from a nearby  convex combination of product states. The algorithm works well for small $N$ and freely available as web applet \cite{statesep}.} 


 \subsection{Completely separable simplex: Classical bits}

The computational basis vectors are pure products, and are the extreme points of a completely separable $(N-1)$-simplex ($N=2^n$). The computational states represent classical bits corresponding to diagonal density matrices:
\be
\rho=\text{diagonal}(\rho_0,\dots,\rho_{N-1}), \quad 1\ge \rho_\alpha\ge0, \quad \sum\rho_\alpha=1
\ee
The simplex is interpreted as the space of probability distributions for {classical  $n$ bits strings}: $\rho_\alpha$ is the probability of the $n$-bits string $\alpha\in \{0,\dots,{N-1}\}$.
\subsection{Entangled {pure} states}\label{s:entangled-pure}

Pure bi-partite states {can be put into equivalence classes labeled by the Schmidt numbers, \cite{nc}, leading to a simple geometric description.} 

Write the bipartite pure state  in $\mathbb{C}^{N_1}\otimes \mathbb{C}^{N_2}$, ($N_1\le N_2$), in the Schmidt decomposition  \cite{nc},
\be\label{e:schmidt}
\ket\psi=\sum_{j=0}^{N_1-1} \sqrt{p_j}\ket{\phi_j}\otimes\ket{\chi_j},\quad \braket{\phi_j}{\phi_k}=\braket{\chi_j}{\chi_k}=\delta_{jk}
\ee
with $p_j\ge 0$ probabilities. The simplex
\be
1\ge p_0\ge \dots\ge p_{N_1-1}> 0, \quad \sum p_j=1
\ee
has the pure product state as the extreme point 
\be
(1,0,\dots, 0)
\ee
All other points of the simplex represent entangled states. The extreme point
\be
\frac 1 {N_1} (1,1,\dots, 1)
\ee
is the maximally entangled state. {Most pure states are entangled}. (In contrast to the density matrix perspective, where by Eq.~(\ref{e:dims}), the separable states are of full dimension.)

We denote by $\ket 0$ the standard maximally entangled state\footnote{We henceforth set $N_1=N_2=M$ to simplify the notation.}:
\be
\ket{\beta_0}=\frac 1{\sqrt {{M}}}\sum_{j=0}^{{{M}}-1} \ket{j}\otimes \ket{j}
\ee
{Let $\sigma_\mu$ be $M^2$ hermitian and mutually orthogonal $M\times M$ matrices i.e.}
\be
\sigma_\mu=\sigma_\mu^*, \quad Tr \sigma_\mu\sigma_\nu= M\delta_{\mu\nu}, \quad \mu\in 0,\dots,M^2-1
\ee
The  projection {$P_0=\ket{\beta_0}\!\bra{\beta_0}$} can be written in terms of $\sigma_\mu$ as: 
\begin{align}\label{e:beta}
{P_0}&=\frac 1{ {{M}}}\sum_{jk=0}^{{{M}}-1} \ket{jj} \bra{kk}  \nonumber\\ 
&=\frac 1 {{{M}}^3}\sum_{jk=0}^{{{M}}-1}\sum_{\mu,\nu=0}^{{{M}}^2-1} \bra{kk} \sigma_\mu\otimes\sigma_\nu^t \ket{jj} \sigma_\mu\otimes\sigma_\nu^t  
\nonumber\\ 
&=\frac 1 {{{M}}^3}\sum_{\mu,\nu=0}^{{{M}}^2-1}  Tr (\sigma_\mu\sigma_\nu) \sigma_\mu\otimes\sigma_\nu^t 
\nonumber\\ 
&=\frac 1 {{{M}}^2}\sum_{\mu=0}^{{{M}}^2-1}   \sigma_\mu\otimes\sigma_\mu^t 
\end{align}
{In the case of qubits and $N=M^2$, a complete set of mutually orthogonal projections on the $N$ maximally entangled states is:}
\be
P_\alpha=\frac 1 {N}\sum_{\mu=0}^{N^2-1}   \sigma_\alpha\sigma_\mu\sigma_\alpha\otimes\sigma_\mu^t, \quad \alpha\in 0,\dots ,N-1 
\ee
{This is a natural generalization of the Bell basis of two qubits, to $2n$ qubits.}

{In the two qubits case, $M=2$, an equal mixture of two Bell states, is a separable state\footnote{{Here $\sigma_\mu$ are the usual Pauli matrices and in particular $\sigma_0$ is the identity}}:}
\begin{align}\label{e:2bell-sep}
   P_\alpha+P_0&=\frac 1 {4}\sum_{\mu=0}^{3}   (\sigma_\alpha\sigma_\mu\sigma_\alpha+\sigma_\mu)\otimes\sigma_\mu^t \nonumber\\
   &=\frac 1 {2}   (\sigma_0\otimes\sigma_0+\sigma_\alpha\otimes\sigma_\alpha^t)
   \nonumber\\
   &=  \frac{\sigma_0+\sigma_\alpha} 2\otimes\frac{\sigma_0+\sigma_\alpha^t} 2+\frac{\sigma_0-\sigma_\alpha} 2\otimes\frac{\sigma_0-\sigma_\alpha^t} 2
\end{align}
{The two terms on the last line  are products of one dimensional projections, and represent together a mixture of pure product states.}
\subsection{Two types of entangled states}
 {Choosing the basis $\sigma_\alpha$ {made with either symmetric real or anti-symmetric imaginary matrices}, makes partial transposition a reflection in the anti-symmetric coordinates}
\begin{equation}
    \big(\sigma_\alpha\otimes \sigma_\beta\big)^{pt}=\sigma_\alpha\otimes \sigma_\beta^t=\pm \sigma_\alpha\otimes \sigma_\beta
    \end{equation}
Partial transposition \cite{peres,nc} distinguishes between two types of entangled states:
\begin{itemize}
    \item $\rho\ge 0 $ while $\rho^{pt}$ is not a positive matrix. 
    \item Both $\rho,\rho^{pt}\ge 0$ but $\rho$ is not separable. 
\end{itemize}
In the case that $\rho$ is a pure state
or\footnote{Simple geometric proofs for two qubits are given in  \cite{myrheim,ak}. }  $N_1N_2\le 6$,  only the first type exists \cite{hhh}. 

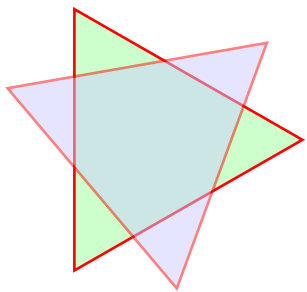
\begin{figure}[ht!]
\centering{} 
\begin{tikzpicture}[scale=2,line width=1pt,black]
   \draw  [fill=green!20!white,draw=red] (-120: 1)--(0:1)--(120:1) -- cycle; 
   \draw  [fill=blue!20!white,draw=red,rotate=40,opacity=.5] (-120: 1)--(0:1)--(120:1) -- cycle; 
  \end{tikzpicture} 
  \caption{The green triangle represent a high dimensional simplex of states and the blue triangle  its partial transposition.  The green triangles that stick out describe entangled states that are discoverable by partial transposition. The  intersection may or may not contain bound entangled states. It is separable iff its vertices are separable. }\label{f:sd}
\end{figure}

The Peres\footnote{Gurvits and Barnum attribute the test to an older 1976 paper of Woronowicz. Apparently, nothing is ever discovered for the first time (M. Berry's law).} {entanglement test \cite{peres} checks  the non-positivity of $\rho^{pt}$ and uncovers entangled state of the first type, (see Fig. \ref{f:sd}). Local operations can not convert states of the second kind into states of the first kind since positivity of partial transposition of $\rho$ implies the positivity under partial transposition of a local operation}
\begin{equation}
   ( M\otimes N)\rho   (M^*\otimes N^*)
\end{equation}
{In particular, one can not distill Bell pairs from entangled states of the second kind by local operations. This is the reason why states of the second kind are  called ``bound entangled":  The Bell pairs used to produce them can not be recovered.}
\subsection{The largest ball of bi-partite separable states}\label{s:ptb}
{The Gurvits-Barnum ball was introduced in  section \ref{s:inscribed} as the largest inscribed ball in $D_N$: }
\begin{equation}\label{e:gb-ball-again}
B_{gb}= \left\{\mathbf{x}\Big||\mathbf{x}|\le r_0^2 \right\}\subseteq D_N
\end{equation}
Since partial transposition is a reflection in the $\sigma_\alpha$ coordinates, and any sphere centered at the origin is invariant under reflection, we have that
\begin{equation}
  B_{gb}\subseteq D_N\cap D_N^{pt}  
\end{equation}
$B_{gb}$ therefore does not contain entangled states that are discoverable by the Peres test.

{Gurvits and Barnum replace partial transposition by contracting positive maps of the form $\id\otimes\phi(D)$ to show \cite{GB},}  that $B_{gb}$ is a ball {of bi-partite separable} states
\begin{equation}
 B_{gb}\subseteq S_{N_1,N_2}   
\end{equation}
\subsection{Entanglement witnesses}\label{s:witness}
An entanglement witness for a given partition, $(N_1,\dots, N_n)$, is a Hermitian matrix $W$ so that  
\begin{equation}\label{e:witness}
Tr (W\rho)\ge 0 \quad \forall\quad  \rho\in S_{N_1,\dots, N_n}    
\end{equation}
This definition makes the set of witnesses a convex cone (see Fig. \ref{f:witness}). 
\begin{rem}
We consider $W\ge 0$ a witness  even though it is  ``dumb" as it does not identify any entangled state. This differs from the definition used in various other places where witnesses are required to be non-trivial, represented by an indefinite $W$.  Non-trivial witnesses have the drawback that they do not form a convex cone.     
\end{rem}

The inequality, Eq~({\ref{e:witness}), is sharp for $\rho$  in the interior of $S_{N_1,\dots N_n}$. As the fully mixed state belongs to the interior }
\begin{equation}
Tr (W)=Tr (W\,\id )> 0   
\end{equation}
we may normalize witnesses to have a unit trace and represent them, alongside the states, by 
\be\label{e:witness-coord}
W(\mathbf{w}) =\frac {\id_N+\sqrt{N-1}\,  \mathbf{w}\cdot \boldsymbol{\sigma}}N,\quad \mathbf{w}\in\mathbb{R}^{N^2-1}
\ee
{We shall show that:}
\begin{equation}
\text{Bi-partite witnesses}\subseteq B_1= \left\{\mathbf{x}\Big||\mathbf{x}|\le 1 \right\}
\end{equation}
{This follows from}
\be\label{e:scalar-prod}
N\,Tr\,(W\rho)=1+(N-1) \mbf{x\cdot w}
\ee 
and
\begin{align}
\text{Bi-partite witnesses}&=
    \left\{\mathbf{w}\Big|\mbf{x\cdot w}\ge -\frac 1 {N-1} \  \forall\   \mathbf{x}\in S_{N_1,N_2}\right\}\nonumber \\
    &\subseteq 
    \left\{\mathbf{w}\Big|\mbf{x\cdot w}\ge -\frac 1 {N-1} \  \forall\   \mathbf{x}\in B_{gb}\right\}\nonumber \\
    &=B_1
\end{align}
\begin{figure}
    \centering
   			\begin{tikzpicture}[scale=6/3,line width=1pt,black]	
  				\draw (0,0) circle (1 cm);
  			\draw[fill=orange!20!white, draw=orange!50!black] (180:1) arc (180:0:1).. controls (0,-.3).. (180:1);	\draw[fill=green!20!white, draw=green!50!black] (180:.2) arc (180:360:.2)--(0,1)--(180:.2);
  				\draw[fill,red] (0,0) circle (.05 cm);
			 \end{tikzpicture}
    \caption{The separable states (green) and the corresponding witnesses, (Schematic).}
    \label{f:witness}
\end{figure}
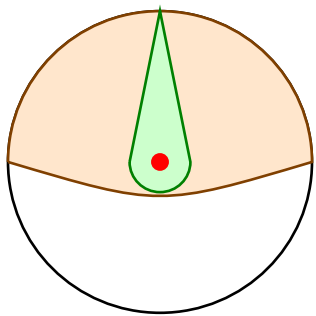
\begin{exa}
 {A witness for the partitioning $N=N_1N_2,\ N_1,N_2\ge M>1$, is:}
   \begin{equation}
      {S}=\sum_{j,k=1}^{M} \ket{j}\bra{k}\otimes\ket{k}\bra{j}
   \end{equation}
$S$ is indeed an entanglement witness since
 \begin{equation}\label{e:swap-c}
      \bra{\psi\otimes\phi}{S}\ket{\psi\otimes\phi}=\left|\sum_{j=1}^{M} \braket{\psi}{j}\braket{j}{\phi}\right|^2
      \ge 0
   \end{equation}
{Since $Tr \,{S}= M $, the corresponding normalized witness is } 
 \begin{equation}\label{e:beta-pt}
      W_{S}=\frac {S} {M}= \big(\ket\beta\bra\beta\big)^{pt}
   \end{equation}
and the equality on  the right follows from {the first line of} Eq.~(\ref{e:beta}).

Since partial transposition is an isometry, it is clear that $W_S$ lies at the same distance from the maximally mixed state as the pure state $\ket{\beta}\bra{\beta}$. In  particular, the associated vector $\mathbf{w}_S$ lies on the unit sphere.
\end{exa}
\begin{rem}
      {In the case that the partitioning is to two isomorphic Hilbert spaces,  $N_1=N_2$, $S$ is the swap. In a coordinate free notation}
       \begin{equation}
      {S}\ket{\psi}\otimes \ket{\psi}=\ket{\phi}\otimes \ket{\phi}
   \end{equation}
  {$\ket\beta$ is then the Bell state.} 
\end{rem}
\subsection{Entangled states and witnesses near the Gurvits-Barnum ball}\label{s:near-bg}

\begin{figure}[ht!]
\begin{center}
			\begin{tikzpicture}[scale=6/3,line width=1pt,black]	
				  \filldraw[fill=green!20!white, draw=green!50!black] (0,0) circle (.3 cm);
  				\draw (0,0) circle (1 cm);
  				\node at (0,1.1) {$\ket{\beta}\!\bra{\beta}$};
  				\node[below] at (0,-.3) {$\rho^{pt}$};
  					\draw[fill,blue] (0,-.35) circle (.03 cm);
  					\node[above] at (0,.35) {$\rho$};
  					\draw[fill,blue] (0,.35) circle (.03 cm);
  					\draw[dashed, orange] (-2,0) --(2,0);
  						\node[below] at (0,-1) {$W_S$};
  							\draw[fill,red] (0,1) circle (.03 cm);
  								\draw[fill,red] (0,-1) circle (.03 cm);
			 \end{tikzpicture}
			 \caption{The figure shows the maximally entangled state $\ket\beta\bra\beta$ on the unit circle, its witness $\rho^{pt}$, the entangled state $\rho$ and the normalized swap witness, $W_S$. $\rho$ and $\rho^{pt}$ lie close to the boundary of the Gurvits-Barnum ball of separable states. The dashed line represent the reflection plane associated with partial transposition. }\label{f:gbk}
			 \end{center}
	  \end{figure}
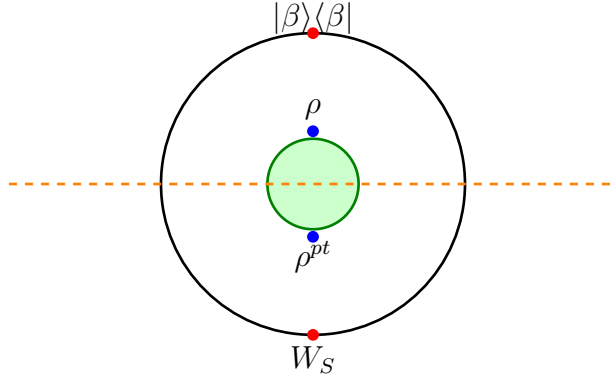

Near the boundary of $B_{gb}$ one can find  entangled states and (non-trivial) entanglement witnesses,  see Fig.~(\ref{f:gbk}).

To see this, consider the bi-partitionning $N=M^2$.  Since $S^2=\id$ and $S=S^*$,
\begin{equation}
P_S=\frac{\id+S}2, \quad P_A=\frac{\id-S}2
\end{equation}
{are orthogonal projections. $P_S$ projects on the states that are symmetric under swap, and $P_A$ on the anti-symmetric ones. }
Hence,
 \begin{equation}
 Tr\, P_A=\frac{M(M-1)}2, \quad Tr\, P_S=\frac{M(M+1)}2
\end{equation}
The state
\begin{equation}
\rho = \frac{1+\varepsilon }{M(M-1)} P_A +
\frac{1-\varepsilon }{M(M+1)} P_S, \quad 0< \varepsilon\le 1
\end{equation}
is entangled with the swap as witness. Indeed,
\begin{align}
Tr\,(S\rho) &= \frac{1+\varepsilon }{M(M-1)} Tr\, (SP_A) +
\frac{1-\varepsilon }{M(M+1)} Tr\, (SP_S)\nonumber \\
 &= -\frac{1+\varepsilon }{M(M-1)} Tr\, (P_A) +
\frac{1-\varepsilon }{M(M+1)} Tr\, (P_S)
\nonumber \\
 &= -\frac{1+\varepsilon }{2} +
\frac{1-\varepsilon }{2}=-\varepsilon
\end{align}
{ When $\varepsilon$ is small, $\rho$ is close to the Gurvits-Barnum ball. One way to see this is to compute its purity}
\begin{align}
 Tr \rho^2&=\left(\frac{1+\varepsilon }{M(M-1)}\right)^2 Tr P_A +
\left(\frac{1-\varepsilon }{M(M+1)}\right)^2 Tr P_S\nonumber \\
&=\frac{M(1+\varepsilon^2)+
2\varepsilon} {M(M^2-1)}
\end{align}
Using  Eq.~(\ref{e:purity}) to translate purity to the radius one finds, after some algebra,
\begin{equation}
 r(\rho)=r_0^2\left(1+\varepsilon \sqrt N\right)
\end{equation}
Since partial transposition is an isometry, $\rho^{pt}$ is also near the Gurvits-Barnum ball. It is an entanglement witness for the Bell state:
\begin{equation}
  -\varepsilon = Tr (\rho S)=Tr (\rho^{pt} S^{pt})=M \bra\beta\rho^{pt} \ket\beta
\end{equation}
and we have used Eq.~(\ref{e:beta-pt}) in the last step.

\subsection{A Clifford ball of separable states }
{The separable states, in $\mathbb{C}^N=\mathbb{C}^M\otimes\mathbb{C}^M $ contain a $2\ell-1$ ball with radius  $1/\sqrt{N-1}$}. 
{$\ell=O(\log M)$, } is the (maximal) number of anti-commuting (generalized) Pauli matrices $\sigma_\mu$, acting on $\mathbb{C}^M$, {(see Eq.~(\ref{e:clifford}))}. We call this ball the Clifford ball.{ It is larger than the}  Gurvits-Barnum ball, {whose radius is $1/(N-1)$}, but it lives in a lower dimension. 

{A standard construction of  $2\ell-1$ anti-commuting Pauli matrices, acting on $\mathbb{C}^N, \ N=M^2$, from $\ell$ anti-commuting matrices acting on $\mathbb{C}^M$ is}
\begin{equation}
    \sigma_\mu\otimes \sigma_\ell, \quad \id\otimes \sigma_\nu,\quad \mu=1\dots \ell,\quad \nu=1\dots \ell-1 
\end{equation}
{Consider the $2\ell-1$ dimensional family of quantum states in  $\mathbb{C}^N$, parametrized by\footnote{Note the change of normalization relative to Eq.~(\ref{e:sigma-coord}).} $\mathbf{a,b}\in\mathbb{R}^\ell$}
\begin{align}\label{e:rho-clifford}
   \rho=\frac{\id_M\otimes \id_M +\mathbf{a} \cdot \boldsymbol{\sigma}\otimes\sigma_\ell+\id\otimes \mathbf{b}\cdot\boldsymbol{\sigma}}{M^2} \ge 0 , 
\end{align}
{where $
\boldsymbol{\sigma}=(\sigma_1,\dots,\sigma_\ell)$ is a vector of $M\times M$ (generalized, anti-commuting) Pauli matrices. We shall show that for $\mathbf{a}^2+\mathbf{b}^2\le 1$, {with $b_\ell=0$,} $\rho$ is essentially equivalent to a family of 2-qubit states, which are manifestly separable. Re-scaling the $\mathbf{a,b}$  coordinates to fit with the convention in Eq.(\ref{e:sigma-coord}) gives the radius {$r_0=1/\sqrt{N-1}$}.}

{Note first that since}
\begin{equation}
   (\mathbf{a} \cdot \boldsymbol{\sigma})^2={\mathbf{a}^2} \id_M, \quad Tr\,  (\mathbf{a} \cdot \boldsymbol{\sigma})=0 
\end{equation}
{we may write}
\begin{equation}
   \mathbf{a} \cdot \boldsymbol{\sigma}={|\mathbf{a}|}\id_{M/2}\otimes  Z  
\end{equation}
\added{}{where $Z$ is the standard $2\times 2$ Pauli matrix. Similarly, since}
\begin{equation}
   \sigma_\ell^2=\id_M,  \quad  (\mathbf{b} \cdot \boldsymbol{\sigma})^2= {{\mathbf{b}}^2} \id_M, \quad \{\sigma_\ell,\mathbf{b} \cdot \boldsymbol{\sigma}\}=0,
\end{equation}
{we may write}
\begin{equation}
  \sigma_\ell=Z \otimes \id_{M/2},  \quad  \mathbf{b} \cdot \boldsymbol{\sigma}= {|\mathbf{b}|}X\otimes \id_{M/2}
\end{equation}
 The numerator in Eq.~(\ref{e:rho-clifford}) takes the form
\begin{equation}
   \id_{M/2}\otimes\Big( \id_2 \otimes \id_2+{|\mathbf{a}|} Z \otimes Z+{|\mathbf{b}|}\id_2\otimes X \Big)\otimes \id_{M/2}
\end{equation}
{The brackets can be written as:}
\begin{equation}
  \ket{0}\!\bra{0}\otimes \left(\id+ {|\mathbf{a}|}Z+{|\mathbf{b}|}X\right)
  +\ket{1}\!\bra{1}\otimes \left(\id- {|\mathbf{a}|}Z+{|\mathbf{b}|}X\right)
\end{equation}
{$\left(\id\pm {|\mathbf{a}|}Z+{|\mathbf{b}|}X\right)\ge0$ when $\mathbf{a}^2+\mathbf{b}^2\le 1$, and the resulting expression for $\rho$  is  manifestly separable. Rescaling the coordinates {to agree with the normalization in Eq.~(\ref{e:sigma-coord})}, gives the radius $r_0=1/\sqrt{N-1}$ for the Clifford ball of separable state. }

\section*{Acknowledgement} The research has been supported by ISF. We thank Yuval Lemberg for  help in the initial stages of this work and Andrzej Kossakowski for his encouragement.
 \newpage
 \appendix

\section{The average purity of quantum states}\label{a:mom}

{In section \ref{s:concentration} we quoted a result of \cite{zs}, Eq.~(\ref{e:zs}), which allows to explicitly compute the radius of inertia $r_e$ of $D_N$ as a rational function of $N$.   The aim of this appendix is to give an elementary derivation of this formula.}

{The measure $d\mu$ on the space of density matrices in  section \ref{s:concentration} is a special case of a more general measures $d\mu_{N,K}$ when $N=K$. The measures $d\mu_{N,K}$ are the induced measure on density matrices acting on $\mathbb{C}^N$, obtained from the uniform measure on pure states on $\mathbb{C}^N\otimes \mathbb{C}^K $ with $K\ge N$,  by partial tracing over the second factor\footnote{In the case $K=N-1$ the measure is concentrated on the boundary of $D_N$ being proportional to $\delta(\det\rho)$.}.  They are all simply related} \cite{zs}
\begin{equation}
    d\mu_{N,K}=\frac 1{Z_{N,K}} \big(\det(\rho)\big)^{K-N} dx_1\dots dx_{N^2-1}, \quad K\ge N
\end{equation}
{$Z_{N,K}$ is a normalization factor.  The Euclidean measure $d\mu$ {manifestly} corresponds to the case $N=K$.  }

{The derivation given below of Eq.~(\ref{e:zs}) is simpler than the original derivation in \cite{zs} in that  it avoids  the constraint associated with the normalization of the wave functions. {This allows to reduce the computation of}  moments of $\mathbf{x}^2$ with  respect to the measure  $d\mu_{N,K}$ to an exercise in Gaussian integration.}  

Let $\braket{\alpha j}{\psi}=\xi_{\alpha j}$ be the amplitudes of the pure state $\ket{\psi}$ in $\mathbb{C}^N\otimes \mathbb{C}^K $. The first factor, {$\alpha$}, is the system and the second,  {$j$}, is the ancila. The density matrix $\rho$ is obtained by  partial tracing the ancila: 
\begin{equation}
    \bra\alpha\rho\ket\beta= \bra\alpha \big(Tr_K \ket{\psi} \bra{\psi}\big)\ket\beta=\sum_j \xi_{\alpha j}\bar \xi_{\beta j}=(\xi \xi^*)_{\alpha\beta}
\end{equation}
{where $\xi$ is an $N\times K$ matrix. The requirement $K\ge N $ guarantees that (generically) $\rho$ is full rank.}

{Choosing $Re\, \xi_{\alpha j}$ and $Im\, \xi_{\alpha j}$ to be normally distributed i.i.d., gives a uniform measure} on pure states,  $d\mu_{\psi}$, which is unitary invariant under $U(NK)$. 
The {induced} measure on the density matrices $d\mu_{N,K}$
{is}:
\begin{equation}
    d\mu_{NK}=d\rho \int \delta\left(\rho-Tr_K\ket{\psi}\!\bra{\psi}\right) d\mu_{\psi}
\end{equation}
Since the Gaussian measure for $\xi$ allows for states that are not normalized, the measure $d\mu_{N,K}$ allows for any $Tr \rho\ge 0$. This means that to compute the moments of normalized density matrices we need to  compute {an integral of a ratio:}
\begin{equation}\label{e:int_ratio}
   \int d\mu_{N,K}\, \frac{Tr\,{(\rho^2)}}{(Tr \, \rho)^2}
\end{equation}
{Remarkably, this reduces to computation of two Gaussian integrals.  
The reason for this }  is that the measure factors {into radial and angular part:}
\begin{equation}
   d\mu_{N,K} =  d\mu_{|\rho|}\otimes d\mu_\Omega
\end{equation}
{Since $Tr \rho=\|\psi\|^2$ is invariant under an arbitrary unitary in $\mathbb{C}^N\otimes\mathbb{C}^K $, and such unitaries act on the measure $d\mu_{N,K}$ as rotations, } $Tr \rho$ depends only on the first, radial, coordinate while $\frac{Tr(\rho^2)}{(Tr\rho)^2}$ depends only on the second, angular, part. {Hence} 
\begin{align}
   \int d\mu_{N,K} Tr\left(\rho^2\right) &=\int  d\mu_{N,K} \frac{Tr(\rho^2)}{(Tr\rho)^2} (Tr\rho)^2\nonumber\\
   &=\int  d\mu_\Omega \frac{Tr(\rho^2)}{(Tr\rho)^2} \int d\mu_{|\rho| } (Tr\rho)^2\nonumber\\
   &=\int  d\mu_{N,K} \frac{Tr(\rho^2)}{(Tr\rho)^2} \int d\mu_{N,K} (Tr\rho)^2
\end{align}
({Assuming $d\mu_{NK}$ normalized.}) The integral in Eq.~(\ref{e:int_ratio}) is  {therefore the ratio of} two Gaussian integrals. 
Wick theorem (for the standard normal distribution) gives   
\begin{align}
   \int d\mu_{N,K} Tr \left(\rho^2\right) &=  \sum_{jk\alpha\beta}\int d\mu_{\psi} \xi_{\alpha j}\bar \xi_{\beta j}\bar\xi_{\beta k}\xi_{\alpha k}
   \nonumber\\
   &=  \sum_{jk\alpha\beta}(\delta_{\alpha\beta}+\delta_{jk})\nonumber \\
   &=NK(K+N)\end{align}
Similarly
\begin{align}
   \int d\mu_{N,K} (Tr\rho)^2 &=  \sum_{jk\alpha\beta}\int d\mu_{G} \xi_{\alpha j}\bar \xi_{\alpha j}\bar\xi_{\beta k}\xi_{\beta k}
   \nonumber\\
   &=  \sum_{jk\alpha\beta}(1+\delta_{jk}\delta_{\alpha\beta})\nonumber \\
   &=NK(NK+1)\end{align}
So finally,
\begin{equation}\label{e:tr-r-2}
   \int d\mu_{N,K}\, \frac{Tr\,{(\rho^2)}}{(Tr \, \rho)^2}=\frac{N+K}{KN+1}
\end{equation}
This reduces to Eq.~(\ref{e:2m}) when $N=K$.

{The computation of higher moments can be similarly reduced to a (tedious) combinatoric problem.}

\section{The N dimensional unit cube is almost a ball}

The fact that $D_N$ looks like a ball in most directions is a general fact about convex bodies in high dimensions. It is instructive to see this happening for the unit cube in $N$ dimensions
  \be
 C=\left\{\mathbf{x}\Big|\mathbf{x}=(x_1,\dots,x_N), \quad  |x_j|\le \frac 12 ,\quad j=1,\dots, N\right\}
  \ee
The radius of inertia of $C$ is
\be\label{e:re}
r_e^2=\average{\mathbf{x}^2}=\sum \average{x_j^2}=N\int_{-1/2}^{1/2}x^2 dx=\frac{N}{12}
\ee
Let us now consider  $r(\theta)$, defined as the maximal $r$  that  is inside the cube for a given direction $\theta$.  

Choose a random direction $\boldsymbol{\theta}=(\theta_1,\dots,\theta_N)$ by  picking $\theta_j$ to be normal iid with 
 \be
 \theta_j\sim {\cal N}\left[0,\frac 1 N\right], \quad x_j=r \theta_j
  \ee
When $N$ is large there is a ``phase transition" in the sense that
\be\label{e:rc}
Prob\big(r \,\boldsymbol{\theta}\in C\big)\approx \begin{cases} 1 & r<r_C\\ 0 & r>r_C
\end{cases}, \quad r_C=\frac 1 2\sqrt{\frac{ N}{\log N}}
  \ee

To see this we first observe that the probability that $x\sim {\mathcal N}\left(0,\sigma^2\right)$ takes values outside the interval $[- x_0,x_0]$ is given by the complementary error function
\begin{align*}
Prob\left(|x|>x_0\right)
=\sqrt{\frac 2{\pi }} \int_{x_0/\sigma}^\infty e^{-x^2/2} dx=\text{erfc}(x_0/\sigma)
\end{align*}
Anticipating the result, {Eq.~(\ref{e:rc})}, let us replace $r$ by its re-scaled version $k$:  
\be
r=   k r_C
\ee
For the case at hand 
\[
\sigma^2=1/N, \quad x_0=\frac 1 {2r}=\frac 1{ 2kr_C}
\]
The probability for {$r|\theta_j|=|x_j|\le 1/2$}  happening for all coordinates simultaneously is
\[
Prob\left(r \,\boldsymbol{\theta} \in C\right)=\left(1-\text{erfc}\left(\frac {\sqrt{\log N} }{k}\right)\right)^N
\]
Since ${\rm erfc}(x)\in[0,1]$ for $x\in[0,\infty)$, the limit $N\to\infty$ tends to 
\be
Prob\left(r \,\theta\in C\right)\to\begin{cases}
 0&  $if $ N\text{erfc}\left(\frac {\sqrt{\log N} }{k}\right)\to\infty\\
 1& $if $ N\text{erfc}\left(\frac {\sqrt{\log N} }{k}\right)\to 0
\end{cases}
\ee
Since
\be
 N\text{erfc}\left(\frac {\sqrt{\log N} }{k}\right)\to\begin{cases}
 \infty&  $for  $ k>1\\
 0& $for  $ k\le 1
\end{cases}
\ee
Eq.~(\ref{e:rc}) follows.



\end{document}